\documentclass{article} 
\usepackage{iclr2025_conference,times}
\usepackage{graphicx}
\usepackage{array}
\usepackage{hyperref}
\usepackage{colortbl}
\usepackage{booktabs}
\usepackage{amsmath}
\usepackage{amssymb}
\usepackage{verbatim}
\usepackage{booktabs}
\usepackage{multirow}
\usepackage{makecell}
\usepackage{multicol}
\usepackage{tcolorbox}
\usepackage{color}
\usepackage{url}
\usepackage{caption}
\definecolor{yan}{rgb}{0.5,0.3,0.5}

\definecolor{ben}{rgb}{0.9,0.,0.5}

\definecolor{navyblue}{RGB}{191, 209, 229} 
\definecolor{light_yellow}{RGB}{255,243,194}
\definecolor{orange}{RGB}{255,200,100}
\definecolor{red}{RGB}{255, 0, 0}
\definecolor{green}{RGB}{0, 176, 80}
\newcommand{\projectname}{ECHOPulse}
\newcommand{\best}{\cellcolor{orange}}
\newcommand{\second}{\cellcolor{light_yellow}}

\title{\projectname: ECG Controlled Echocardiograms Video Generation}

\author{\textbf{Yiwei Li}$^{1,2}$\thanks{Equal contribution} , \textbf{Sekeun Kim}$^2$\footnotemark[1] ,\textbf{ Zihao Wu}$^{1,2}$, \textbf{Hanqi Jiang}$^1$, \textbf{Yi Pan}$^1$\\
\textbf{Pengfei Jin}$^2$, \textbf{Sifan Song}$^2$, \textbf{Yucheng Shi}$^1$,  \\
\textbf{Tianming Liu}$^1$\thanks{Corresponding author email: xli60@mgh.harvard.edu} , \textbf{Quanzheng Li}$^2$\footnotemark[2] , \textbf{Xiang Li}$^2$\footnotemark[2]\\
${}^1$ University of Georgia \\
${}^2$Massachusetts General Hospital and Harvard Medical School\\
}

\begin{document}

\maketitle

\begin{abstract}

Echocardiography (ECHO) is essential for cardiac assessments, but its video quality and interpretation heavily relies on manual expertise, leading to inconsistent results from clinical and portable devices. ECHO video generation offers a solution by improving automated monitoring through synthetic data and generating high-quality videos from routine health data. However, existing models often face high computational costs, slow inference, and rely on complex conditional prompts that require experts' annotations. To address these challenges, we propose \projectname{}, an ECG-conditioned ECHO video generation model. \projectname{} introduces two key advancements: (1) it accelerates ECHO video generation by leveraging VQ-VAE tokenization and masked visual token modeling for fast decoding, and (2) it conditions on readily accessible ECG signals, which are highly coherent with ECHO videos, bypassing complex conditional prompts. To the best of our knowledge, this is the first work to use time-series prompts like ECG signals for ECHO video generation. \projectname{} not only enables controllable synthetic ECHO data generation but also provides updated cardiac function information for disease monitoring and prediction beyond ECG alone. Evaluations on three public and private datasets demonstrate state-of-the-art performance in ECHO video generation across both qualitative and quantitative measures. Additionally, \projectname{} can be easily generalized to other modality generation tasks, such as cardiac MRI, fMRI, and 3D CT generation. Demo can seen from \url{https://github.com/levyisthebest/ECHOPulse_Prelease}.
\end{abstract}

\section{Introduction}

Echocardiography (ECHO) is widely used for evaluating heart function and structure due to its real-time, non-invasive, and radiation-free nature~\citep{sf2009recommendations,otto2013textbook,omar2016advances}. Recent advancements in deep learning have shown promising performance in computer-aided cardiac diagnosis~\citep{oktay2018attention,smistad2020real,kim2021automatic}. However, echocardiography depends on experienced operators, leading to insufficient data for training robust models. Moreover, annotating ECHO data demands expert knowledge, making the process time-consuming and sometimes unavailable~\citep{leclerc2019deep,lu2021lv}.

One promising avenue is the use of video generation models to synthesize ECHO videos. In general domains, text-to-video (T2V) models~\citep{yu2023magvit,zhang2024moonshot,wang2024videocomposer,Phenaki} have been successful in creating realistic videos from text prompts. However, applying these models to the medical field remains challenging due to the necessity for precise control over anatomical structures, which cannot vary freely as in general domain. Previous controllable ECHO generation frameworks~\citep{ashrafian2024vision,zhou2024heartbeat,yu2024explainable} rely on carefully curated conditional prompts such as segmentation masks, clinical text, or optical flow, which only experts can provide and fail to incorporate time-series information. Moreover, these diffusion-based models suffer from high computational costs and slow inference. Therefore, a fast and controllable ECHO generation framework without the need for expert prior conditions is highly desirable~\citep{reynaud2024echonet}.

Inspired by these motivations, we propose \projectname{}, an ECHO video generation model conditioned on electrocardiogram (ECG) signals. ECG data are readily accessible in both clinical and outpatient settings~\citep{al20182017}, and even from personal wearable devices~\citep{liu2013wearable}. Since ECGs are typically the first step in evaluating cardiac issues before performing further echocardiography~\citep{al20182017,del2022advances}, they serve as a natural and intrinsic prior for coherent ECHO video generation. By leveraging ECG signals, we can generate ECHO videos without the need for complex conditional prompts or expert annotations. To the best of our knowledge, this is the first work to utilize time-series prompt for ECHO video generation. In clinical scenarios where prior ECHO images exist, \projectname{} can be co-conditioned on both ECG and prior imaging to provide updated cardiac function information, such as left ventricular ejection fraction (LVEF/EF) for cardiac disease monitoring and prediction~\citep{ponikowski20162016}, enhancing computer-aided diagnosis beyond ECG alone.

Our main contributions are:
\begin{itemize}
    \item We propose \projectname{}, a fast and controllable ECHO video generation framework conditioned on ECG for the first time, which also shows potential to generalize to other modality generation tasks, such as cardiac MRI, fMRI, and 3D CT generation.
    \item By utilizing readily available ECG data for controllable and precise ECHO generation personalized for each patient, we bypass the need for paired expert annotations, showing potential for scaling to larger video datasets.
    \item Evaluations on two public and one private dataset demonstrate that our framework achieves SOTA performance in both quantitative and qualitative metrics.
    \item We collect a synthetic ECHO video dataset consisting of ECHO videos paired with ECG signals and segmentation masks, which we plan to release to enable researchers to evaluate the practicality and reliability of synthetic ECHO video generation.
\end{itemize}


\section{RELATED WORK}
\subsection{Visual Tokenization}
Visual Tokenization is crucial for converting images or videos into discrete tokens that can be processed by language models. One of the foundational approaches to visual tokenization is VQ-VAE \citep{VanDenOord2017NeuralDiscrete}, which uses a convolutional neural network (CNN) to encode visual data, followed by a vector quantization (VQ) step. This method assigns each embedding to the closest entry in a codebook, converting continuous embeddings into discrete tokens. Video tokenization presents greater challenges, and VQGAN has been adapted to address these difficulties \citep{Phenaki, yu2023magvit}. The current state-of-the-art in video tokenization is MAGVIT2~\citep{Yu2024LanguageModelBeats}, which employs an enhanced 3D architecture, leverages an inflation technique for initialization based on image pre-training, and incorporates robust training losses to improve performance.

\subsection{Video Generation}




Recent video generation advances have increasingly focused on discrete representations to manage the high dimensionality of video data. Models like VQ-VAE~\citep{VanDenOord2017NeuralDiscrete} convert continuous video frames into discrete vector indices, reducing data complexity and enabling more efficient sequence modeling. Phenaki~\citep{Phenaki} uses a causal Transformer with discrete temporal embeddings for variable-length video generation, while Masked Generative Transformer~\citep{maskgit} leverages discrete techniques for sparse spatiotemporal relationships. These studies highlight the advantages of discrete representations in balancing complexity and efficiency, especially for long sequences. Moreover, autoregressive models with discrete tokens~\citep{yu2022generating} generate diverse, high-quality content, and discrete diffusion models~\citep{austin2021structured} extend the diffusion process to the discrete symbol space, enhancing scalability and long-range video modeling.

\subsection{ECHO Video Generation}
Ultrasound imaging is a non-invasive, real-time, and portable medical imaging modality that plays a crucial role in clinical diagnostics. However, the quality and interpretation of ultrasound images heavily depend on the operator's expertise, and obtaining high-quality ultrasound datasets is challenging. Consequently, the generation of ultrasound images has attracted considerable attention in recent years. Recent works, such as HeartBeat \citep{zhou2024heartbeat}, have leveraged deep learning techniques like diffusion models \citep{ho2020denoising} to achieve controllable synthesis of echocardiograms. By incorporating multimodal conditions—such as cardiac anatomical annotations and functional indices—and employing diffusion models, these methods can generate realistic ultrasound videos with specific cardiac morphology and function. This approach enriches training datasets and enhances algorithm robustness. \citet{reynaud2024echonet} introduced the EchoNet-Dynamic model, which utilizes spatiotemporal three-dimensional convolutional neural networks to segment the left ventricle frame by frame and estimate the ejection fraction (EF), outperforming manual measurements. Such advancements hold promise for assisting clinicians in more precise and efficient cardiac function assessments. These techniques typically use images as conditional inputs to generate corresponding ultrasound images. Recent studies have also explored generating ultrasound video sequences from single-frame images \citep{reynaud2023feature}, offering new possibilities for dynamic ultrasound applications like cardiac function evaluation. However, these studies are conditioned on carefully curated conditional prompts such as segmentation masks and clinical text but do not leverage time-series ECG data. In contrast, our work is the first to efficiently generate controllable ECHO videos conditioned on ECG data.

\section{METHODS}
\label{Methods}

\begin{figure}[t]

\begin{center}
\includegraphics[width=\textwidth]{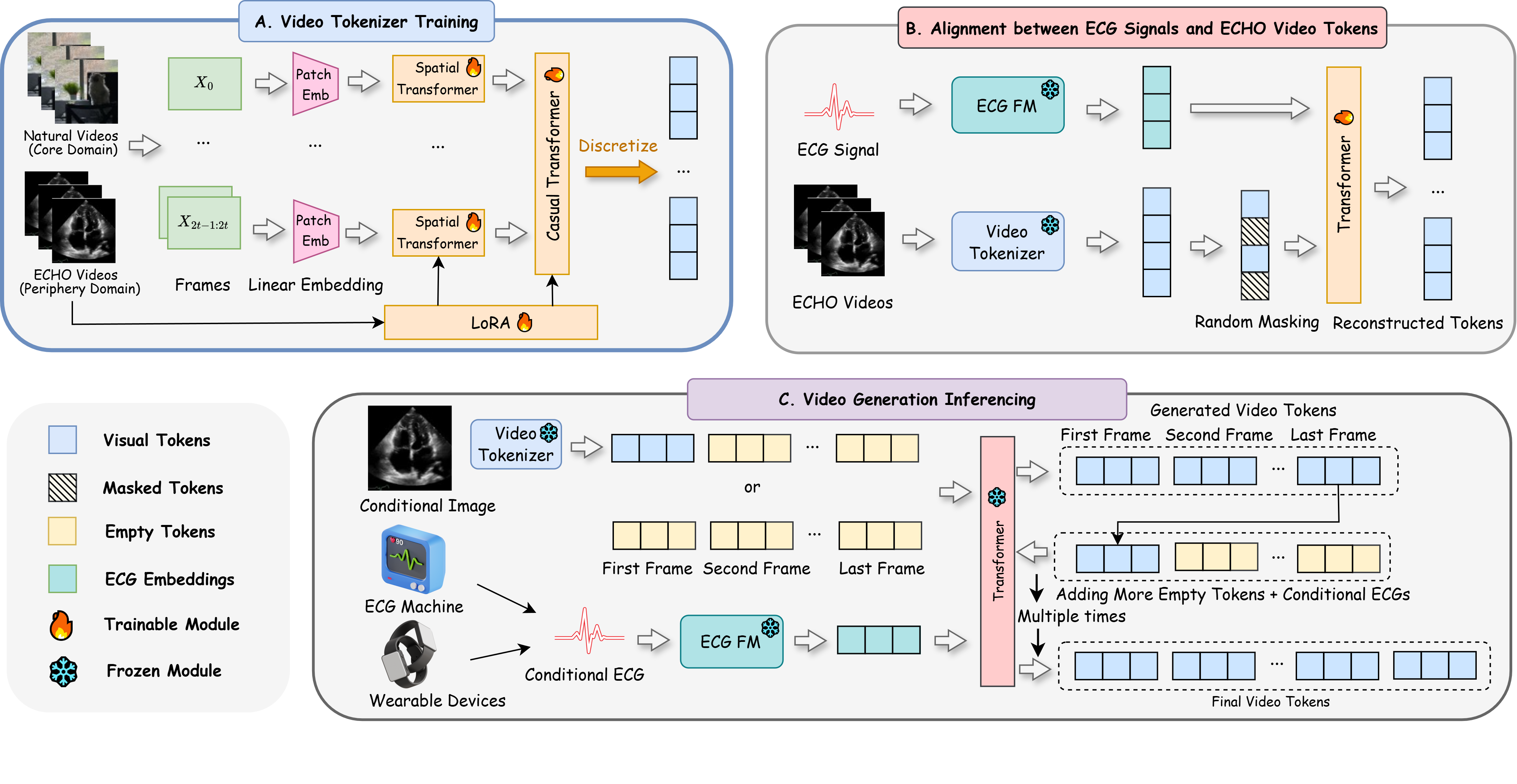}
\end{center}
\caption{The pipeline of the \projectname{}.
 \projectname{} contains a two-step training procedure. a) The first step trains the video tokenizer on the natural video dataset first and fine-tunes it on the public ECHO video dataset. b) The second step trains the transformer via the input token produced by ECG foundation model and video tokenizer, pretrained in the first step. Followed by the video generation procedure (c), \projectname{} accepts input with or without a conditional image. The empty tokens will be reconstructed via the frozen transformer, trained in the second step, through the guidance of ECG siganls. \projectname{} is capable of generating continuous long videos by sequentially shifting the token sequence and integrating new ECG inputs.}
 \label{pipeline}
 \vspace{-0.9cm}
\end{figure}

\projectname{}, as shown in Figure \ref{pipeline}, consists of three key components: (1) a video tokenization model that encodes videos into temporal token sequences, (2) a masked generative transformer that learns the latent alignment between video tokens and ECG signal, and (3) a progressive video generation pipeline that enables fast, unlimited video generation. The video tokenization and transformer components are trained separately.

\subsection{Video Tokenization}
\label{Video encoder}
\begin{figure}[t]
\begin{center}
\includegraphics[width=1\textwidth]{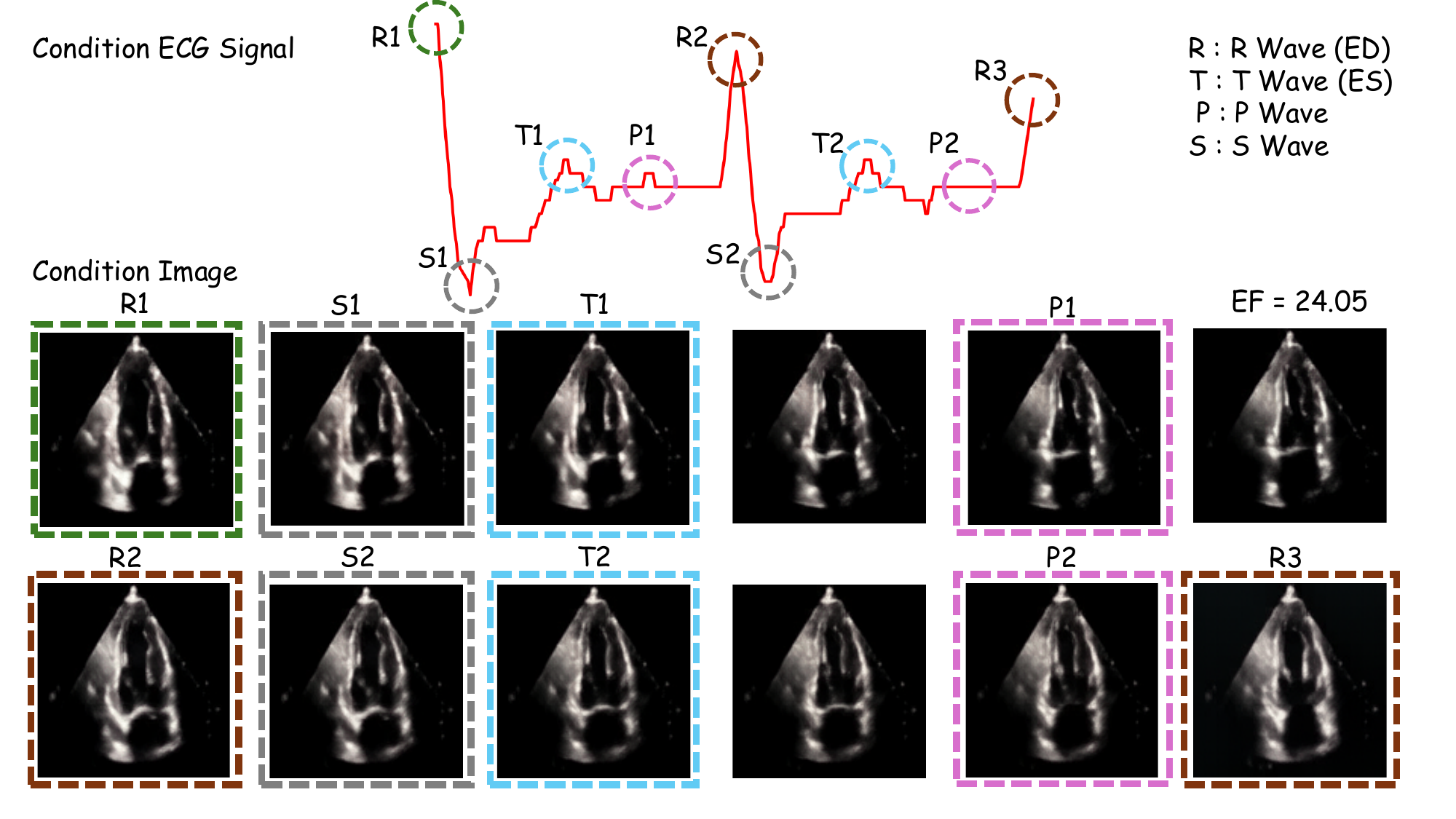}
\end{center}
\caption{Video generation example of \projectname. The inputs to ECHOPulse consist of a conditioning image and an ECG signal. Utilizing these inputs as constraints, \projectname{} generates corresponding videos. The RSTP waves represent the four phases of the ECG. The R wave corresponds to end-diastole (ED), which is the frame with the largest ventricular segmentation area, while the T wave corresponds to end-systole (ES), the frame with the smallest ventricular area. In this case the EF of the generated ECHO video is 24.05.}
\vspace{-0.5cm}
\label{fig:result_2}
\end{figure}

The key to generating videos from ECG signals lies in capturing the temporal correlation between the video and ECG data. Unlike text~\citep{touvron2023llama}, ECG is a continuous temporal signal with variable sequence lengths. As shown in Figure \ref{fig:result_2}, each ECG phase should correspond to different stages of the ECHO. Thus, encoding the entire video into a single embedding risks disrupting temporal alignment between ECG and video, also complicating autoregressive video generation and reducing generalizability. To temporally align ECHO video with the various phases of the ECG and support autoregressive video generation, an effective approach is to discretize the video into tokens while splitting the ECG into patches and embedding them for cross-modal alignment. Drawing inspiration from ViViT~\citep{arnab2021vivit} and C-ViViT~\citep{Phenaki}, we developed \projectname{}'s video tokenization model to meet these specific requirements.

\paragraph{Model architecture:}
The encoding starts with a video sequence of $T+1$ frames, each with resolution $H \times W$ and $C$ channels, denoted as $\mathbf{X} \in \mathbb{R}^{(T+1) \times H \times W \times C}$. The sequence is compressed into tokens of size $(T' + 1) \times H' \times W'$, where the first frame is tokenized independently, and subsequent frames are spatio-temporal tokens that autoregressive depending on the previous. Non-overlapping patches of size $P_h \times P_w \times C$ (for images) and $P_t \times P_h \times P_w \times C$ (for video) are extracted, resulting in $T' = \frac{T}{P_t}$, $H' = \frac{H}{P_h}$, and $W' = \frac{W}{P_w}$. Each patch is flattened and projected into a $D$-dimensional space, forming a tensor of shape $(T' + 1) \times (H' \times W') \times D$. Spatial transformers are applied with self attention, followed by temporal transformers with causal attention. The patch embeddings $\mathbf{Z} \in \mathbb{R}^{(T'+1) \times H' \times W' \times D}$ are then quantized into codewords $\mathbf{C_z}$ via vector quantization. The decoder reverses the process by transforming quantized tokens into embeddings, applying temporal and spatial transformers, and mapping the tokens back to pixel space via linear projection to reconstruct the frames $\hat{\mathbf{X}}$.

Additionally, Our framework incorporates the Lookup-Free Quantization (LFQ) method, inspired by the advancements in MAGVIT2~\citep{Yu2024LanguageModelBeats}, to streamline the quantization process. Unlike traditional vector quantization that relies on an explicit codebook lookup, LFQ parameterizes the quantization operation directly. Moreover, due to the lack of high-quality medical video datasets comparable in scale to natural video datasets, we introduce LoRA (Low-Rank Adaptation)~\citep{hu2021lora} to enhance the learning efficiency and generalizability for medical video generation. This also lays the foundation for offering personalized video generation services for users or patients in future applications.

\paragraph{Loss design:} During video tokenization training phrase, the total loss function $\mathcal{L}_{\text{total}}$ is formulated as:
\begin{equation}
    \mathcal{L}_{\text{total}} = \mathcal{L}_{\text{recon}} + \mathcal{L}_{\text{percep}} + \mathcal{L}_{\text{VQ}} + \lambda_{\text{adaptive}} \cdot \mathcal{L}_{\text{GAN}},
\end{equation}
where $\mathcal{L}_{\text{recon}}$ is the reconstruction loss, $\mathcal{L}_{\text{percep}}$ is the perceptual loss, $\mathcal{L}_{\text{VQ}}$ represents the vector quantization loss, and $\mathcal{L}_{\text{GAN}}$ is the generator adversarial loss. The parameter $\lambda_{\text{adaptive}}$ dynamically adjusts the contribution of the adversarial loss relative to the perceptual loss during training.

Specifically, the reconstruction loss $\mathcal{L}_{\text{recon}}$ is computed using the Mean Squared Error (MSE) to ensure the reconstructed video $\hat{\mathbf{V}}$ to closely match the original video $\mathbf{V}$ at the pixel level:
\begin{equation}
    \mathcal{L}_{\text{recon}} = \frac{1}{N} \left\| \mathbf{V} - \hat{\mathbf{V}} \right\|_2^2,
\end{equation}
where $N$ is the total number of pixels across all frames, $\mathbf{V}_i$ and $\hat{\mathbf{V}}_i$ are the $i$-th pixel of the original video and the reconstructed video, respectively.

To improve the perceptual quality of the reconstructed video, perceptual loss $\mathcal{L}_{\text{percep}}$ is employed to measure the similarity~\citep{johnson2016perceptual} between the original and reconstructed videos in the feature space of a pre-trained VGG16~\citep{qassim2018compressed} network $\phi$. By extracting high-level features from randomly selected video frames $\mathbf{V}_{f_j}$ and $\hat{\mathbf{V}}_{f_j}$, the perceptual loss is defined as:
\begin{equation}
\mathcal{L}_{\text{percep}} = \frac{1}{M} \sum_{j=1}^{M} \left\| \phi(\mathbf{V}_{f_j}) - \phi(\hat{\mathbf{V}}_{f_j}) \right\|_2^2,
\end{equation}
where $M$ is the number of feature elements, and $\phi(\cdot)$ denotes the VGG feature extraction operation. 

The vector quantization loss~\citep{van2017neural} $\mathcal{L}_{\text{VQ}}$ regularizes the encoder outputs by ensuring that they are close to their nearest codebook entries, encouraging efficient and discrete representation learning. The VQ loss is defined as:
\begin{equation}
    \mathcal{L}_{\text{VQ}} = \beta \cdot \left\| \text{sg}[\mathbf{z}_e] - \mathbf{e} \right\|_2^2,
\end{equation}
where $\mathbf{z}_e$ is the encoder output, $\mathbf{e}$ is the closest codebook entry, and $\beta$ is a weighting parameter. The term $\text{sg}[\cdot]$ represents the stop-gradient operation, which ensures that the encoder is encouraged to commit to a discrete codebook entry without backpropagating gradients through this operation. 

The generator adversarial loss $\mathcal{L}_{\text{GAN}}$ drives the model to produce videos that are indistinguishable from real videos, as evaluated by a discriminator $D$. We employ the hinge loss formulation for stable adversarial training:
\begin{equation}
    \mathcal{L}_{\text{GAN}} = -\mathbb{E}_{\hat{\mathbf{V}} \sim P_G} \left[ D(\hat{\mathbf{H}}) \right],
\end{equation}
where $\hat{\mathbf{H}}$ is a video generated by the model, sampled from the generator’s distribution $P_G$, and $D(\hat{\mathbf{H}})$ is the discriminator's output for the generated video. 

To balance the adversarial loss $\mathcal{L}_{\text{GAN}}$ and the perceptual loss $\mathcal{L}_{\text{percep}}$, we introduce an adaptive weight $\lambda_{\text{adaptive}}$. This weight is computed dynamically based on the gradients of both losses with respect to the parameters of the decoder’s last layer~\citep{johnson2016perceptual}. The adaptive weight is calculated as:
\begin{equation}
    \lambda_{\text{adaptive}} = \frac{\left\| \nabla_{\theta} \mathcal{L}_{\text{percep}} \right\|_2}{\left\| \nabla_{\theta} \mathcal{L}_{\text{GAN}} \right\|_2 + \epsilon},
\end{equation}
where $\theta$ are the parameters of the decoder’s last layer, $\nabla_{\theta} \mathcal{L}_{\text{percep}}$ and $\nabla_{\theta} \mathcal{L}_{\text{GAN}}$ represent the gradients of the perceptual and adversarial losses with respect to $\theta$, and $\epsilon$ is a small constant to avoid division by zero. 

\subsection{Alignment between Video and ECG Tokens }
Next, we embed the ECG signals and align them with the video tokens. To fully leverage the time-series ECG data, we implemented the following encoding and alignment design.
\paragraph{ECG encoder:}To better couple with echo video, the time-series ECG is first divided into non-overlapping temporal and spatial patches, which are then linearly projected into a high-dimensional space~\citep{na2024guiding}. Self-attention mechanisms capture spatio-temporal relationships across time segments, and a masked autoencoder (MAE)~\citep{he2022masked} further refines the model's understanding of ECG dynamics, transforming them into separate embedding features. In this work, we divide ECG signals into patches using the embedding layer of an ECG foundation model (ECG-FM), whose strong predictive performance ensures the effectiveness of this patch-based representation. The proposed patchify approach is as follows:
The raw ECG signal, denoted as $\mathbf{X} \in \mathbb{R}^{L \times T}$, where $L$ is the number of leads and $T$ is the time length, is split into non-overlapping spatio-temporal patches. Each lead $l$ produces patches $\mathbf{P}^l \in \mathbb{R}^{n \times p}$, where $n = \frac{T}{p}$ and $p$ is the patch size.

Each patch $\mathbf{P}_i^l$ is passed through the ECG-FM encoder, producing a $D$-dimensional embedding:
\[
\mathbf{Z}_i^l = \text{ECG-FM}(\mathbf{P}_i^l) \in \mathbb{R}^D,
\]
which results in a sequence of embeddings $\mathbf{Z} = \{\mathbf{Z}_1^l, \mathbf{Z}_2^l, \dots \} \in \mathbb{R}^{L \times n \times D}$.
\label{VideoGeneration}
\paragraph{Aligning ECG and video token sequences for video prediction:} Our video generation framework employs a bidirectional transformer with MVTM~\citep{maskgit} to predict video tokens. We mask tokens during training and use parallel decoding during inference to iteratively fill in missing tokens based on context from all directions. This design allows for faster and more efficient video synthesis, leveraging the bidirectional self-attention mechanism to condition video generation on both past and future frames. The masked token prediction is driven by minimizing the negative log-likelihood of the masked tokens:
\begin{equation}
    \mathcal{L}_{\text{mask}} = - \mathbb{E}_{\mathbf{X}} \left[ \sum_{i=1}^{N} m_i \log P(\mathbf{y}_i | \mathbf{Y}_M) \right],
\end{equation}
where \( \mathbf{y}_i \) are the masked tokens, \( m_i \) indicates the mask for each token, and \( \mathbf{Y}_M \) represents the partially masked token sequence.

At each iteration, the mask scheduling function \( \gamma(t/T) \) determines the fraction of tokens to mask:
\begin{equation}
    \gamma(t/T) = \cos\left(\frac{\pi t}{2 T}\right),
\end{equation}
where \( t \) is the current iteration and \( T \) is the total number of iterations. The cosine schedule progressively reduces the mask ratio, enabling a gradual refinement of the generated video~\citep{maskgit}.

The bidirectional transformer architecture allows for flexible video generation guided by various conditioning inputs. We also adapted T5X~\citep{roberts2023scaling} following~\citep{reynaud2024echonet,ashrafian2024vision}'s text-based ECHO video generation method to create a text-to-ECHO video version for comparative experiments. What's more, we introduce an ECG mask similar to a text mask. This mask ensures that certain portions of the ECG signal are masked out during training, forcing the model to learn to generate corresponding video frames based on the visible ECG features. 





\paragraph{Optimization:} To ensure the transformer model fully learns the entire action of ECHO videos, we add a critic loss function~\citep{lezama2022improved}. The critic takes a sequence of predicted tokens $\hat{\mathbf{y}}$ and real tokens $\mathbf{y}$ and computes whether the predicted tokens are real or fake. 
The critic's objective is expressed as:
\begin{equation}
    \mathcal{L}_{\text{critic}} = -\mathbb{E}_{\mathbf{y}} \left[ \log D(\mathbf{y}) \right] - \mathbb{E}_{\hat{\mathbf{y}}} \left[ \log(1 - D(\hat{\mathbf{y}})) \right],
\end{equation}
where $D(\mathbf{y})$ is the critic's prediction for the real tokens, and $D(\hat{\mathbf{y}})$ is its prediction for the fake tokens (predicted by transformer). In our framework, $\mathbf{y}$ represents the actual video tokens, and $\hat{\mathbf{y}}$ are the predicted tokens sampled during the transformer reconstruction process.



In practice, the critic’s loss function is computed using the following binary cross-entropy (BCE) loss~\citep{goodfellow2016deep}:
\begin{equation}
    \mathcal{L}_{\text{BCE}} = - \frac{1}{N} \sum_{i=1}^{N} \left[ y_i \log(\hat{y_i}) + (1 - y_i) \log(1 - \hat{y_i}) \right],
\end{equation}
where $y_i$ are the true labels (1 for real tokens, 0 for fake tokens) and $\hat{y_i}$ are the critic’s predicted probabilities.
\subsection{Video Generation}
We first train a video tokenization model and an ECG-conditioned video token prediction transformer. To generate videos conditioned on ECG, we pass the empty video value through the video tokenization model to obtain empty tokens and use the ECG tokens as input to the pre-trained transformer. This mirrors the training setup where all video tokens are masked, enabling video prediction from ECG. Additionally, as the first frame is independently tokenized, using it as a condition mimics the training scenario where only the first frame is visible, allowing for video generation based on the initial frame. Similarly, by conditioning on the final segment of a previous video and another ECG inputs, the transformer can generate coherent subsequent frames.
\paragraph{Progressive video generation with auto-regressive extrapolation:} 
During inference, video tokens are sampled iteratively, similar to the process described in MaskGIT~\citep{maskgit}, using classifier-free guidance with a scaling factor $\lambda$ to balance the alignment between the generated video and the provided ECG condition. Once the first $t_x + 1$ frames are generated in the latent space, we can further extrapolate additional frames in an auto-regressive manner by re-encoding the last $K$ generated frames using the video encoder. These tokens are then used to initialize the pre-trained transformer, which generates the remaining video tokens conditioned on the provided ECG. This approach enables rapid and length-unconstrained generation of ECHO videos, significantly enhancing its clinical applicability.




\section{EXPERIMENTS}
\label{Experiments}


\begin{figure}[t]
\begin{center}
\includegraphics[width=1\textwidth]{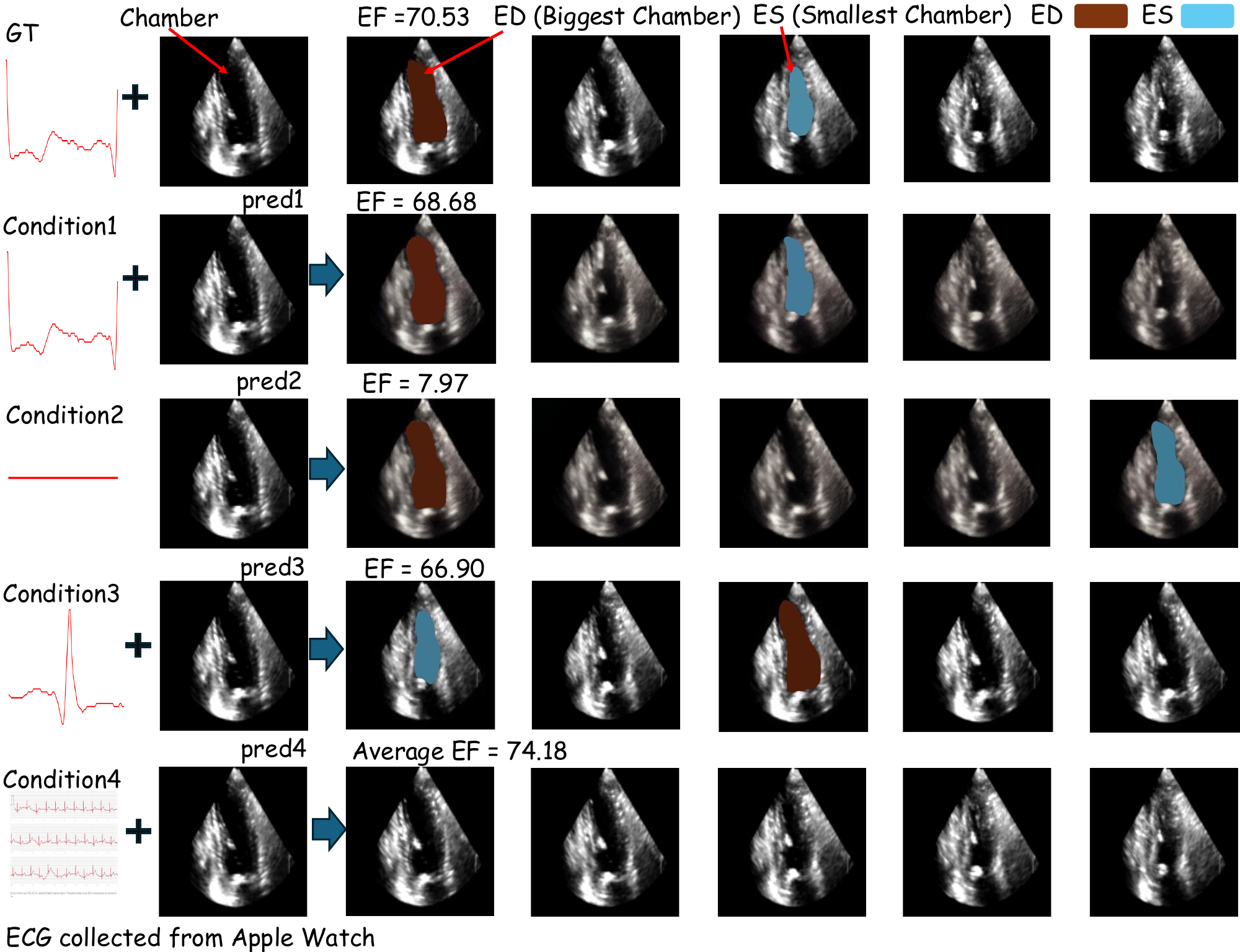}
\end{center}
\caption{Video generation results from \projectname{} under different ECG conditions. The first row displays the ground truth, while the second row shows the generated video using the same ECG. The third row illustrates results under flat ECG conditions. The fourth condition depicts a shifted R wave, and the final condition, collected from an Apple Watch, represents the most common scenario. The ECG for Condition 4 was collected directly from the Health app on Apple Watch v9.}
\vspace{-0.2cm}
\label{fig:result_3}
\end{figure}

\paragraph{Dataset and implementations:} In this experiment, the process is conducted in a distributed manner. Initially, we train the video tokenization model on the Webvim-10M dataset~\citep{Bain2021FrozenInTime}, followed by fine-tuning on the CAMUS dataset~\citep{Leclerc2019DeepLearningSegmentation} to adapt the tokenization for medical imaging tasks. With the introduction of LFQ, the codebook size is set to \( 2^{13} \). All natural videos are resized to 128x128 resolution with 11 frames. The temporal patch size is set to 2, and the spatial patch size is set to 8, resulting in a token sequence with a length of 1536. More detailed information can be seen in Appendix Table \ref{tab:model_hyperparameters} and Table \ref{tab:training_hyperparameters}.
 
For the subsequent training of \projectname{}, we utilized a private dataset consisting of 94,078 video and ECG samples. A total of 473 videos were extracted from the dataset to serve as the test dataset, and all video generation evaluations were performed on this test dataset. During the training process, all ECHO videos are resized to a format of 128×128 with 11 frames in the dataloader to match the input requirements of the video encoder. Due to the large volume and diverse sources of our dataset, the overall quality is not as consistent as that of publicly available ECHO datasets. Therefore, each video undergoes random contrast augmentation during loading to enhance its visual quality. We used the pretrained weight of ST-MEM~\citep{na2024guiding} as ECG signal encoder. Similarly, the evaluation metrics were maintained to ensure a fair and direct comparison. This allowed us to evaluate the model's performance and validate the effectiveness of the proposed method. 

For evaluation, we use the CAMUS~\citep{leclerc2019deep} and EchoNet-Dynamic~\citep{ouyang2020video} datasets. CAMUS, a public dataset with 500 ECHO videos with diverse spatial resolutions, is currently the highest-quality public ECHO dataset available. EchoNet-Dynamic includes 10,030 4-chamber cardiac ultrasound sequences with a resolution of 112×112 pixels.

All experiments were conducted on 8 NVIDIA H-100 GPUs, each with 80GB of memory. Due to potential privacy concerns, the private dataset used in this study cannot be made publicly available. However, we have released all the code, scripts, and the pretrained ECG2Video model weights to facilitate further research and replication of our work.
\subsection{Video Tokenization Reconstruction Performance}

In our experiments, we conduct a quantitative comparison between the performance of the diffusion-based EchoNet-Synthetic model and our VQ-VAE-based model, primarily evaluated using MSE (Mean Squared Error)~\citep{gauss1809theoria} and MAE (Mean Absolute Error)~\citep{qi2020mean} on reconstruction results of different datasets. As shown in Table~\ref{tabel:rec results}.

Specifically, the EchoNet-Synthetic model achieves an MSE of \(3.00 \times 10^{-3}\) and an MAE of \(3.00 \times 10^{-2}\) on the Echo-Dynamic dataset. In comparison, our VQ-VAE based model, initially trained on a natural video dataset, exhibits an MSE of \(5.02 \times 10^{-3}\) and an MAE of \(4.08 \times 10^{-2}\) on the same dataset. Although the initial errors are slightly higher, after only one epoch of fine-tuning, the MSE of our model significantly improves to \(2.89 \times 10^{-3}\), with the MAE reduced to \(2.92 \times 10^{-2}\), surpassing the performance of the diffusion-based model.


%
On the private dataset, the results followed a similar trend. Before fine-tuning, the VQ-VAE model had an MSE of \(6.09 \times 10^{-3}\) and an MAE of \(4.41 \times 10^{-2}\). However, after one epoch of domain transfer, the MSE was significantly reduced to \(2.91 \times 10^{-3}\), and the MAE to \(2.97 \times 10^{-2}\).

\begin{table}[t]
\centering
    \captionsetup{type=table}
    \captionof{table}{\textbf{Reconstruction metrics:} Videos reconstructed using video tokenization and diffusion-based models are evaluated across two datasets with MAE and MSE. The \colorbox{orange}{best} results are highlighted.}
    \resizebox{\linewidth}{!}{%
    \begin{tabular}{lccc}
        \toprule
         \textbf{Methods}& \textbf{Dataset} & \bf MSE$\downarrow$ & \bf MAE$\downarrow$ \\ \midrule
         EchoNet-Synthetic \citep{reynaud2024echonet} & Echo-Dynamic & \( 3.00 \times 10^{-3} \) & \(  3.00 \times 10^{-2} \)\\ 
        \projectname{} (Only natural videos) & Echo-Dynamic & \( 5.02 \times 10^{-3} \) & \(  4.08 \times 10^{-2} \)\\ 
        \projectname{} (Domain transfer) & Echo-Dynamic & \(\best{2.89 \times 10^{-3}} \) & \( \best{2.92 \times 10^{-2}} \) \\ \midrule
        \projectname{} (Only natural videos) & Private data & \( 6.09 \times 10^{-3} \) & \(  4.41 \times 10^{-2} \)\\ 
        \projectname{} (Domain transfer) & Private data & \(\best{2.91 \times 10^{-3}} \) & \(\best{2.97\times 10^{-2}} \) \\
        \bottomrule
    \end{tabular}
    }
\label{tabel:rec results}
\end{table}

These results indicate that, despite the common expectation that diffusion-based models perform better in terms of reconstruction quality, our VQ-VAE model was able to quickly converge in a new domain. With only one epoch of fine-tuning, it achieved results comparable to or even surpassing those of the diffusion-based model. This further demonstrates the potential of our model for clinical applications, where it can quickly adapt to new tasks or datasets without the need for retraining from scratch, offering greater practical utility and flexibility.

\subsection{Video Generation Performance}

We evaluate our model on two objectives:
1) image quality (FID~\citep{heusel2017gans}, FVD~\citep{unterthiner2018towards}, SSIM~\citep{wang2004image}) and left ventricular ejection fraction (LVEF/EF) accuracy. Since we are the first to explore ECG-guided ECHO video generation, and most prior works generating ECHO videos from text have not been open-sourced, we were unable to reproduce their results on our dataset. However, those related papers have clearly reported experimental parameters, datasets, and data split methods. Therefore, we trained \projectname{} using T5X~\citep{roberts2023scaling} as the text encoder on both the CAMUS and EchoNet datasets. All experimental settings, including video resolution and parameters, were kept consistent with those reported in previous studies. 

\begin{table}[t]
\caption{\textbf{Image quality metrics:} The videos generated by different methods under various conditions across three datasets are evaluated using three image quality metrics. The \colorbox{orange}{best} and \colorbox{light_yellow}{second} results are highlighted. Note that EchoNet-Synthetic~\citep{reynaud2024echonet} did not report the SSIM metric and did not specify whether the results correspond to A2C or A4C views.}
\centering
\resizebox{\linewidth}{!}{
\begin{tabular}{lccccccccc}
\toprule
\multirow{2}{*}{\bf Methods} & \multirow{2}{*}{\bf Dataset} & \multirow{2}{*}{\bf Condition} & \multicolumn{3}{c}{\bf A2C} & \multicolumn{3}{c}{\bf A4C} \\
\cmidrule(lr){4-6} \cmidrule(lr){7-9}
 &  &  & \bf FID$\downarrow$ & \bf FVD$\downarrow$ & \bf SSIM$\uparrow$ & \bf FID$\downarrow$ & \bf FVD$\downarrow$ & \bf SSIM$\uparrow$ \\
\midrule
MoonShot~\citep{zhang2024moonshot} & CAMUS & Text & 48.44 & 202.41 & 0.63 & 61.57 & 290.08 & 0.62 \\
VideoComposer~\citep{wang2024videocomposer} & CAMUS & Text  & 37.68 & 164.96 & 0.60 & 35.04 & 180.32 & 0.61 \\
HeartBeat~\citep{zhou2024heartbeat} & CAMUS & Text & 107.66 & 305.12 & 0.53 & 76.46 & 381.28 & 0.53 \\
\projectname{} (Only natural videos) & CAMUS & Text & 12.71 & 273.15 & 0.61 & 15.38 & 336.04 & 0.58 \\
\projectname{} (Domain transfer) & CAMUS & Text & 5.65 & 211.85 & 0.79 & 8.17 & 283.32 & 0.75 \\
\midrule
EchoDiffusion~\citep{reynaud2023feature} & Echo-Dynamic & Text+EF &  24.00 & 228.00 & 0.48 & 24.00 & 228.00 & 0.48 \\
\projectname{} (Only natural videos) & Echo-Dynamic & Text & 36.10 & 319.25 & 0.39 & 44.21 & 334.95 & 0.35 \\
\projectname{} (Domain transfer) & Echo-Dynamic & Text & 27.50 & 249.46 & 0.46 & 29.83 & 312.31 & 0.41 \\
\midrule
\projectname{} (Only natural videos) & Private data & Text & 27.49 & 291.67 & 0.53 & 34.13 & 374.92 & 0.51 \\
\projectname (Domain transfer) & Private data & Text & 25.44 & 224.90 & 0.54 & 31.21 & 334.09 & 0.54 \\
\projectname{} (Only natural videos) & Private data & ECG & \second{18.73} & \second{200.45} & \second{0.56} & \second{27.37} & \second{302.89} & \second{0.55} \\
\projectname{} (Domain transfer) & Private data & ECG & \best{15.50} & \best{82.44} & \best{0.67} & \best{20.82} & \best{107.40} & \best{0.66} \\
\midrule
HeartBeat~\citep{zhou2024heartbeat} & CAMUS & Text+Sketch+Mask+...  & 25.23 & 97.28 & 0.66 & 31.99 & 159.36 & 0.65 \\
EchoNet-Synthetic~\citep{reynaud2024echonet} & Echo-Dynamic & Video+EF & 16.90 & 87.40 & - & 16.90 & 87.40 & - \\
\bottomrule
\end{tabular}
}
\label{tab:results}
\end{table}

Based on Table \ref{tab:results}, \projectname{} outperforms other models across several key metrics. In the ECG-conditioned setting, it achieves the lowest FID scores of 15.50 (A2C) and 20.82 (A4C), indicating highly realistic video generation. Its FVD scores of 82.44 (A2C) and 107.40 (A4C) further highlight its ability to capture temporal dynamics. Additionally, \projectname{}'s SSIM scores of 0.67 (A2C) and 0.66 (A4C) are higher than those of competing models, showcasing its superior video quality and structural coherence.

\projectname{}+Text significantly outperforms the HeartBeat model, which also uses text control, on the CAMUS dataset, achieving lower FID 5.65/8.17 and higher SSIM scores 0.79/0.75. As CAMUS is currently the highest-quality publicly available ECHO dataset, all models perform better on CAMUS, while the modest FVD scores 211.85/283.32 highlight the limitations of text-based control. \projectname{}+Txt also exceeds EchoDiffusion on EchoNet-Dynamic. While VQ-VAE models may lag behind diffusion in video quality, their compatibility with ECG foundation models, coupled with the strong ECG-ECHO correlation, drives \projectname{}'s superior performance.

After LoRA fine-tuning, \projectname{} shows further improvement but still falls slightly short of HeartBeat, which uses six conditions, and EchoNet-Synthetic, which modifies entire videos based on a given LVEF value. However, \projectname{}+ECG excels on the larger, more complex private dataset, demonstrating the effectiveness of ECG as a condition for video generation.



To validate that the generated videos are synchronized with the ECG signals, we compared the cardiac phases between the generated and original videos, focusing on the end-diastole (ED) and end-systole (ES) phases. ED, which corresponds to the R wave in the ECG, represents the moment when the chamber area is at its largest. Similarly, ES corresponds to the T wave, when the chamber area is at its smallest. As shown in Figure \ref{fig:result_2}, the generated videos, conditioned on the input image and ECG, strictly follow the phase changes dictated by the ECG. A more quantitative evaluation is performed by assessing the LVEF, demonstrating the alignment of cardiac dynamics between the generated and original ECHO video.

To evaluate our model's performance on LVEF accuracy, we applied SAM2~\citep{ravi2024sam} to the generated videos for direct segmentation of the endocardium, epicardium, and left atrium. After segmenting the cardiac chamber structures, we compared the LVEF between the original echocardiograms and the generated ones. $LVEF = \frac{LVED - LVES}{LVED}$. 
This comparison ensured that the generated videos were controlled solely by the ECG signals—introducing temporal information without affecting the structural integrity of the heart images. As shown in Figure \ref{fig:result_3}.

It can be observed that \projectname{} is capable of generating realistic videos with relatively accurate EF values, regardless of whether the input is an original ECG or a flat ECG. In particular, Condition 4 demonstrates the model's ability to generate videos using daily ECG signals collected directly from an Apple Watch. This highlights \projectname{}'s zero-shot capability in handling general ECG signals, showcasing its potential for broad applicability with real-world ECG data.

\begin{table}[t]
\vspace{-0.1cm}
\centering
\caption{\textbf{Clinical and time-inference metrics:} This table presents the EF (Ejection Fraction) differences between the generated videos and the target ones. The sampling time refers to the time required to generate 64 video frames. The \colorbox{orange}{best} results are highlighted.}
\resizebox{\linewidth}{!}{%
\begin{tabular}{@{}lcccccc@{}}
\toprule
\textbf{Methods} & \textbf{Condition} & $\bf R^2$ $\uparrow$ & \textbf{MAE} $\downarrow$ & \textbf{RMSE} $\downarrow$ & \textbf{S. time}$\downarrow$ & \textbf{Parameters}\\ \midrule
EchoDiffusion \citep{reynaud2023feature}  & Image+EF & \best{0.89} & 4.81 & 6.69  & Gen. 146s    &381M\\ 
\projectname{}     & Image+ECG & 0.85 & \best{2.51} & \best{2.86} & \best{Gen. 6.4s} &279M\\
\bottomrule
\end{tabular}
}
\vspace{-0.1cm}
\label{tabel:EF results}
\end{table}

Quantitative results of $LVEF$ are provided in Table \ref{tabel:EF results}, where our model outperforms the other~\citep{reynaud2023feature} in terms of MAE and RMSE. Although the R-square value is slightly lower, this is understandable given the higher difficulty of our task. Additionally, \projectname{} achieves these results with fewer parameters and requires less sampling time to generate higher-quality videos. This not only underscores the validity of using ECG as a conditioning signal but also demonstrates the clinical potential of our model.

\section{CONCLUSION}
In this paper, we present \projectname{}, an innovative framework for generating ECHO videos conditioned on ECG signals, marking a significant advancement in the integration of time-series data for video synthesis. By utilizing VQVAE-based tokenization and a transformer with MVTM-driven generation, \projectname{} not only achieves superior performance in video quality but also addresses key challenge of ejection fraction estimation in cardiac video generation. Our extensive evaluation on both public and private datasets highlights the model's robust potential for enhancing clinical workflows and enabling real-time cardiac monitoring. This work lays the foundation for further exploration in automated, data-driven cardiac assessment, with implications for both research and practical applications.
\vspace{-0.2cm}
\section{LIMITATION AND FUTURE WORK}
\vspace{-0.2cm}
There are several limitations of this study that have yet to be addressed. Although we have expanded the dataset through the generation of synthetic data, we are still unable to automatically ensure that the generated videos are free from potential personal privacy information, which limits our ability to fully release the whole private dataset. Additionally, while \projectname{} can generate videos of unlimited length, it currently lacks the ability to improve video frame rates (FPS) and resolution. In the future, integrating. \projectname{}, when combined with diffusion models and techniques like super-resolution and denoising, could further enhance the quality of generated videos. This approach addresses current limitations and pushes the boundaries of ECHO video generation.







\bibliography{iclr2025_conference}

\begin{thebibliography}{47}
\providecommand{\natexlab}[1]{#1}
\providecommand{\url}[1]{\texttt{#1}}
\expandafter\ifx\csname urlstyle\endcsname\relax
  \providecommand{\doi}[1]{doi: #1}\else
  \providecommand{\doi}{doi: \begingroup \urlstyle{rm}\Url}\fi

\bibitem[Al-Khatib et~al.(2018)Al-Khatib, Stevenson, Ackerman, Bryant, Callans, Curtis, Deal, Dickfeld, Field, Fonarow, et~al.]{al20182017}
Sana~M Al-Khatib, William~G Stevenson, Michael~J Ackerman, William~J Bryant, David~J Callans, Anne~B Curtis, Barbara~J Deal, Timm Dickfeld, Michael~E Field, Gregg~C Fonarow, et~al.
\newblock 2017 aha/acc/hrs guideline for management of patients with ventricular arrhythmias and the prevention of sudden cardiac death: a report of the american college of cardiology/american heart association task force on clinical practice guidelines and the heart rhythm society.
\newblock \emph{Journal of the American College of Cardiology}, 72\penalty0 (14):\penalty0 e91--e220, 2018.

\bibitem[Arnab et~al.(2021)Arnab, Dehghani, Heigold, Sun, Lu{\v{c}}i{\'c}, and Schmid]{arnab2021vivit}
Anurag Arnab, Mostafa Dehghani, Georg Heigold, Chen Sun, Mario Lu{\v{c}}i{\'c}, and Cordelia Schmid.
\newblock Vivit: A video vision transformer.
\newblock In \emph{Proceedings of the IEEE/CVF international conference on computer vision}, pp.\  6836--6846, 2021.

\bibitem[Ashrafian et~al.(2024)Ashrafian, Yazdani, Heidari, Shahriari, and Hacihaliloglu]{ashrafian2024vision}
Pooria Ashrafian, Milad Yazdani, Moein Heidari, Dena Shahriari, and Ilker Hacihaliloglu.
\newblock Vision-language synthetic data enhances echocardiography downstream tasks.
\newblock \emph{arXiv preprint arXiv:2403.19880}, 2024.

\bibitem[Austin et~al.(2021)Austin, Johnson, Ho, Tarlow, and van~den Berg]{austin2021structured}
Jacob Austin, Daniel~D Johnson, Jonathan Ho, Daniel Tarlow, and Rianne van~den Berg.
\newblock Structured denoising diffusion models in discrete state-spaces.
\newblock \emph{Advances in Neural Information Processing Systems}, 34:\penalty0 17981--17993, 2021.

\bibitem[Bain et~al.(2021)Bain, Nagrani, Varol, and Zisserman]{Bain2021FrozenInTime}
Max Bain, Arsha Nagrani, G{\"u}l Varol, and Andrew Zisserman.
\newblock Frozen in time: A joint video and image encoder for end-to-end retrieval.
\newblock In \emph{IEEE International Conference on Computer Vision}, 2021.

\bibitem[Chang et~al.(2022)Chang, Zhang, Jiang, et~al.]{maskgit}
Huiwen Chang, Han Zhang, Lu~Jiang, et~al.
\newblock Maskgit: Masked generative image transformer.
\newblock In \emph{Proceedings of the IEEE/CVF Conference on Computer Vision and Pattern Recognition}, pp.\  11315--11325, 2022.

\bibitem[Del~Torto et~al.(2022)Del~Torto, Guaricci, Pomarico, Guglielmo, Fusini, Monitillo, Santoro, Vannini, Rossi, Muscogiuri, et~al.]{del2022advances}
Alberico Del~Torto, Andrea~Igoren Guaricci, Francesca Pomarico, Marco Guglielmo, Laura Fusini, Francesco Monitillo, Daniela Santoro, Monica Vannini, Alexia Rossi, Giuseppe Muscogiuri, et~al.
\newblock Advances in multimodality cardiovascular imaging in the diagnosis of heart failure with preserved ejection fraction.
\newblock \emph{Frontiers in Cardiovascular Medicine}, 9:\penalty0 758975, 2022.

\bibitem[Gauss(1809)]{gauss1809theoria}
Carl Gauss.
\newblock Theoria motus corporum coelestium in sectionibus conicis solem mabientium [theory of the motion of the heavenly bodies moving about the sun in conic sections].
\newblock \emph{Perthes and Besser, Hamburg, Germany}, 1809.

\bibitem[Goodfellow(2016)]{goodfellow2016deep}
Ian Goodfellow.
\newblock Deep learning, 2016.

\bibitem[He et~al.(2022)He, Chen, Xie, Li, Doll{\'a}r, and Girshick]{he2022masked}
Kaiming He, Xinlei Chen, Saining Xie, Yanghao Li, Piotr Doll{\'a}r, and Ross Girshick.
\newblock Masked autoencoders are scalable vision learners.
\newblock In \emph{Proceedings of the IEEE/CVF conference on computer vision and pattern recognition}, pp.\  16000--16009, 2022.

\bibitem[Heusel et~al.(2017)Heusel, Ramsauer, Unterthiner, Nessler, and Hochreiter]{heusel2017gans}
Martin Heusel, Hubert Ramsauer, Thomas Unterthiner, Bernhard Nessler, and Sepp Hochreiter.
\newblock Gans trained by a two time-scale update rule converge to a local nash equilibrium.
\newblock \emph{Advances in neural information processing systems}, 30, 2017.

\bibitem[Ho et~al.(2020)Ho, Jain, and Abbeel]{ho2020denoising}
Jonathan Ho, Ajay Jain, and Pieter Abbeel.
\newblock Denoising diffusion probabilistic models.
\newblock \emph{Advances in Neural Information Processing Systems}, 33:\penalty0 6840--6851, 2020.

\bibitem[Hu et~al.(2021)Hu, Shen, Wallis, Allen-Zhu, Li, Wang, Wang, and Chen]{hu2021lora}
Edward~J Hu, Yelong Shen, Phillip Wallis, Zeyuan Allen-Zhu, Yuanzhi Li, Shean Wang, Lu~Wang, and Weizhu Chen.
\newblock Lora: Low-rank adaptation of large language models.
\newblock \emph{arXiv preprint arXiv:2106.09685}, 2021.

\bibitem[Johnson et~al.(2016)Johnson, Alahi, and Fei-Fei]{johnson2016perceptual}
Justin Johnson, Alexandre Alahi, and Li~Fei-Fei.
\newblock Perceptual losses for real-time style transfer and super-resolution.
\newblock In \emph{Computer Vision--ECCV 2016: 14th European Conference, Amsterdam, The Netherlands, October 11-14, 2016, Proceedings, Part II 14}, pp.\  694--711. Springer, 2016.

\bibitem[Kim et~al.(2021)Kim, Hedayat, Vaitkus, Belohlavek, Krishnamurthy, and Borazjani]{kim2021automatic}
Taeouk Kim, Mohammadali Hedayat, Veronica~V Vaitkus, Marek Belohlavek, Vinayak Krishnamurthy, and Iman Borazjani.
\newblock Automatic segmentation of the left ventricle in echocardiographic images using convolutional neural networks.
\newblock \emph{Quantitative Imaging in Medicine and Surgery}, 11\penalty0 (5):\penalty0 1763, 2021.

\bibitem[Leclerc et~al.(2019{\natexlab{a}})Leclerc, Smistad, Pedrosa, {\O}stvik, Cervenansky, Espinosa, Espeland, Berg, Jodoin, Grenier, et~al.]{leclerc2019deep}
Sarah Leclerc, Erik Smistad, Joao Pedrosa, Andreas {\O}stvik, Frederic Cervenansky, Florian Espinosa, Torvald Espeland, Erik Andreas~Rye Berg, Pierre-Marc Jodoin, Thomas Grenier, et~al.
\newblock Deep learning for segmentation using an open large-scale dataset in 2d echocardiography.
\newblock \emph{IEEE transactions on medical imaging}, 38\penalty0 (9):\penalty0 2198--2210, 2019{\natexlab{a}}.

\bibitem[Leclerc et~al.(2019{\natexlab{b}})Leclerc, Smistad, Pedrosa, Ostvik, et~al.]{Leclerc2019DeepLearningSegmentation}
S{\'e}bastien Leclerc, Erik Smistad, Jorge Pedrosa, Andre Ostvik, et~al.
\newblock Deep learning for segmentation using an open large-scale dataset in 2d echocardiography.
\newblock \emph{IEEE Transactions on Medical Imaging}, 38\penalty0 (9):\penalty0 2198--2210, Sept. 2019{\natexlab{b}}.

\bibitem[Lezama et~al.(2022)Lezama, Chang, Jiang, and Essa]{lezama2022improved}
Jos{\'e} Lezama, Huiwen Chang, Lu~Jiang, and Irfan Essa.
\newblock Improved masked image generation with token-critic.
\newblock In \emph{European Conference on Computer Vision}, pp.\  70--86. Springer, 2022.

\bibitem[Liu et~al.(2013)Liu, Xie, Zhou, Zou, and Wu]{liu2013wearable}
Jun Liu, Fei Xie, Yaqi Zhou, Qian Zou, and Jianfeng Wu.
\newblock A wearable health monitoring system with multi-parameters.
\newblock In \emph{2013 6th International Conference on Biomedical Engineering and Informatics}, pp.\  332--336. IEEE, 2013.

\bibitem[Lu et~al.(2021)Lu, Chen, Gao, Sun, Ntziachristos, and Li]{lu2021lv}
Tong Lu, Tingting Chen, Feng Gao, Biao Sun, Vasilis Ntziachristos, and Jiao Li.
\newblock Lv-gan: A deep learning approach for limited-view optoacoustic imaging based on hybrid datasets.
\newblock \emph{Journal of biophotonics}, 14\penalty0 (2):\penalty0 e202000325, 2021.

\bibitem[Na et~al.(2024)Na, Park, Tae, and Joo]{na2024guiding}
Yeongyeon Na, Minje Park, Yunwon Tae, and Sunghoon Joo.
\newblock Guiding masked representation learning to capture spatio-temporal relationship of electrocardiogram.
\newblock \emph{arXiv preprint arXiv:2402.09450}, 2024.

\bibitem[Oktay et~al.(2018)Oktay, Schlemper, Folgoc, Lee, Heinrich, Misawa, Mori, McDonagh, Hammerla, Kainz, et~al.]{oktay2018attention}
Ozan Oktay, Jo~Schlemper, Loic~Le Folgoc, Matthew Lee, Mattias Heinrich, Kazunari Misawa, Kensaku Mori, Steven McDonagh, Nils~Y Hammerla, Bernhard Kainz, et~al.
\newblock Attention u-net: Learning where to look for the pancreas.
\newblock \emph{arXiv preprint arXiv:1804.03999}, 2018.

\bibitem[Omar et~al.(2016)Omar, Bansal, and Sengupta]{omar2016advances}
Alaa Mabrouk~Salem Omar, Manish Bansal, and Partho~P Sengupta.
\newblock Advances in echocardiographic imaging in heart failure with reduced and preserved ejection fraction.
\newblock \emph{Circulation research}, 119\penalty0 (2):\penalty0 357--374, 2016.

\bibitem[Otto(2013)]{otto2013textbook}
Catherine~M Otto.
\newblock \emph{Textbook of clinical echocardiography}.
\newblock Elsevier Health Sciences, 2013.

\bibitem[Ouyang et~al.(2020)Ouyang, He, Ghorbani, Yuan, Ebinger, Langlotz, Heidenreich, Harrington, Liang, Ashley, et~al.]{ouyang2020video}
David Ouyang, Bryan He, Amirata Ghorbani, Neal Yuan, Joseph Ebinger, Curtis~P Langlotz, Paul~A Heidenreich, Robert~A Harrington, David~H Liang, Euan~A Ashley, et~al.
\newblock Video-based ai for beat-to-beat assessment of cardiac function.
\newblock \emph{Nature}, 580\penalty0 (7802):\penalty0 252--256, 2020.

\bibitem[Ponikowski et~al.(2016)Ponikowski, Voors, Anker, Bueno, Cleland, Coats, Falk, Gonz{\'a}lez-Juanatey, Harjola, Jankowska, et~al.]{ponikowski20162016}
Piotr Ponikowski, Adriaan~A Voors, Stefan~D Anker, H{\'e}ctor Bueno, John~GF Cleland, Andrew~JS Coats, Volkmar Falk, Jos{\'e}~Ram{\'o}n Gonz{\'a}lez-Juanatey, Veli-Pekka Harjola, Ewa~A Jankowska, et~al.
\newblock 2016 esc guidelines for the diagnosis and treatment of acute and chronic heart failure.
\newblock \emph{Kardiologia Polska (Polish Heart Journal)}, 74\penalty0 (10):\penalty0 1037--1147, 2016.

\bibitem[Qassim et~al.(2018)Qassim, Verma, and Feinzimer]{qassim2018compressed}
Hussam Qassim, Abhishek Verma, and David Feinzimer.
\newblock Compressed residual-vgg16 cnn model for big data places image recognition.
\newblock In \emph{2018 IEEE 8th annual computing and communication workshop and conference (CCWC)}, pp.\  169--175. IEEE, 2018.

\bibitem[Qi et~al.(2020)Qi, Du, Siniscalchi, Ma, and Lee]{qi2020mean}
Jun Qi, Jun Du, Sabato~Marco Siniscalchi, Xiaoli Ma, and Chin-Hui Lee.
\newblock On mean absolute error for deep neural network based vector-to-vector regression.
\newblock \emph{IEEE Signal Processing Letters}, 27:\penalty0 1485--1489, 2020.

\bibitem[Ravi et~al.(2024)Ravi, Gabeur, Hu, Hu, Ryali, Ma, Khedr, R{\"a}dle, Rolland, Gustafson, et~al.]{ravi2024sam}
Nikhila Ravi, Valentin Gabeur, Yuan-Ting Hu, Ronghang Hu, Chaitanya Ryali, Tengyu Ma, Haitham Khedr, Roman R{\"a}dle, Chloe Rolland, Laura Gustafson, et~al.
\newblock Sam 2: Segment anything in images and videos.
\newblock \emph{arXiv preprint arXiv:2408.00714}, 2024.

\bibitem[Reynaud et~al.(2023)Reynaud, Qiao, Dombrowski, Day, Razavi, Gomez, Leeson, and Kainz]{reynaud2023feature}
Hadrien Reynaud, Mengyun Qiao, Mischa Dombrowski, Thomas Day, Reza Razavi, Alberto Gomez, Paul Leeson, and Bernhard Kainz.
\newblock Feature-conditioned cascaded video diffusion models for precise echocardiogram synthesis.
\newblock In \emph{International Conference on Medical Image Computing and Computer-Assisted Intervention}, pp.\  142--152. Springer, 2023.

\bibitem[Reynaud et~al.(2024)Reynaud, Meng, Dombrowski, Ghosh, Day, Gomez, Leeson, and Kainz]{reynaud2024echonet}
Hadrien Reynaud, Qingjie Meng, Mischa Dombrowski, Arijit Ghosh, Thomas Day, Alberto Gomez, Paul Leeson, and Bernhard Kainz.
\newblock Echonet-synthetic: Privacy-preserving video generation for safe medical data sharing.
\newblock \emph{arXiv preprint arXiv:2406.00808}, 2024.

\bibitem[Roberts et~al.(2023)Roberts, Chung, Mishra, Levskaya, Bradbury, Andor, Narang, Lester, Gaffney, Mohiuddin, et~al.]{roberts2023scaling}
Adam Roberts, Hyung~Won Chung, Gaurav Mishra, Anselm Levskaya, James Bradbury, Daniel Andor, Sharan Narang, Brian Lester, Colin Gaffney, Afroz Mohiuddin, et~al.
\newblock Scaling up models and data with t5x and seqio.
\newblock \emph{Journal of Machine Learning Research}, 24\penalty0 (377):\penalty0 1--8, 2023.

\bibitem[SF(2009)]{sf2009recommendations}
NAGUEH SF.
\newblock Recommendations for the evaluation of left ventricular diastolic function by echocardiography.
\newblock \emph{J Am Soc Echocardiogr}, 22:\penalty0 107--133, 2009.

\bibitem[Smistad et~al.(2020)Smistad, {\O}stvik, Salte, Melichova, Nguyen, Haugaa, Brunvand, Edvardsen, Leclerc, Bernard, et~al.]{smistad2020real}
Erik Smistad, Andreas {\O}stvik, Ivar~Mj{\aa}land Salte, Daniela Melichova, Thuy~Mi Nguyen, Kristina Haugaa, Harald Brunvand, Thor Edvardsen, Sarah Leclerc, Olivier Bernard, et~al.
\newblock Real-time automatic ejection fraction and foreshortening detection using deep learning.
\newblock \emph{IEEE transactions on ultrasonics, ferroelectrics, and frequency control}, 67\penalty0 (12):\penalty0 2595--2604, 2020.

\bibitem[Touvron et~al.(2023)Touvron, Martin, Stone, Albert, Almahairi, Babaei, Bashlykov, Batra, Bhargava, Bhosale, et~al.]{touvron2023llama}
Hugo Touvron, Louis Martin, Kevin Stone, Peter Albert, Amjad Almahairi, Yasmine Babaei, Nikolay Bashlykov, Soumya Batra, Prajjwal Bhargava, Shruti Bhosale, et~al.
\newblock Llama 2: Open foundation and fine-tuned chat models.
\newblock \emph{arXiv preprint arXiv:2307.09288}, 2023.

\bibitem[Unterthiner et~al.(2018)Unterthiner, Van~Steenkiste, Kurach, Marinier, Michalski, and Gelly]{unterthiner2018towards}
Thomas Unterthiner, Sjoerd Van~Steenkiste, Karol Kurach, Raphael Marinier, Marcin Michalski, and Sylvain Gelly.
\newblock Towards accurate generative models of video: A new metric \& challenges.
\newblock \emph{arXiv preprint arXiv:1812.01717}, 2018.

\bibitem[Van Den~Oord \& Vinyals(2017)Van Den~Oord and Vinyals]{VanDenOord2017NeuralDiscrete}
Aaron Van Den~Oord and Oriol Vinyals.
\newblock Neural discrete representation learning.
\newblock \emph{Advances in Neural Information Processing Systems}, 30, 2017.

\bibitem[van~den Oord et~al.(2017)van~den Oord, Vinyals, and Kavukcuoglu]{van2017neural}
Aaron van~den Oord, Oriol Vinyals, and Koray Kavukcuoglu.
\newblock Neural discrete representation learning.
\newblock \emph{Advances in neural information processing systems}, 30, 2017.

\bibitem[Villegas et~al.(2022)Villegas, Babaeizadeh, Kindermans, et~al.]{Phenaki}
Ruben Villegas, Mohammad Babaeizadeh, Pieter-Jan Kindermans, et~al.
\newblock Phenaki: Variable length video generation from open domain textual descriptions.
\newblock In \emph{International Conference on Learning Representations}, 2022.

\bibitem[Wang et~al.(2024)Wang, Yuan, Zhang, Chen, Wang, Zhang, Shen, Zhao, and Zhou]{wang2024videocomposer}
Xiang Wang, Hangjie Yuan, Shiwei Zhang, Dayou Chen, Jiuniu Wang, Yingya Zhang, Yujun Shen, Deli Zhao, and Jingren Zhou.
\newblock Videocomposer: Compositional video synthesis with motion controllability.
\newblock \emph{Advances in Neural Information Processing Systems}, 36, 2024.

\bibitem[Wang et~al.(2004)Wang, Bovik, Sheikh, and Simoncelli]{wang2004image}
Zhou Wang, Alan~C Bovik, Hamid~R Sheikh, and Eero~P Simoncelli.
\newblock Image quality assessment: from error visibility to structural similarity.
\newblock \emph{IEEE transactions on image processing}, 13\penalty0 (4):\penalty0 600--612, 2004.

\bibitem[Yu et~al.(2024{\natexlab{a}})Yu, Chen, Zhou, Chen, Duan, Huang, Zhou, Tao, Yang, and Ni]{yu2024explainable}
Junxuan Yu, Rusi Chen, Yongsong Zhou, Yanlin Chen, Yaofei Duan, Yuhao Huang, Han Zhou, Tan Tao, Xin Yang, and Dong Ni.
\newblock Explainable and controllable motion curve guided cardiac ultrasound video generation.
\newblock \emph{arXiv preprint arXiv:2407.21490}, 2024{\natexlab{a}}.

\bibitem[Yu et~al.(2023)Yu, Cheng, Sohn, Lezama, Zhang, Chang, Hauptmann, Yang, Hao, Essa, et~al.]{yu2023magvit}
Lijun Yu, Yong Cheng, Kihyuk Sohn, Jos{\'e} Lezama, Han Zhang, Huiwen Chang, Alexander~G Hauptmann, Ming-Hsuan Yang, Yuan Hao, Irfan Essa, et~al.
\newblock Magvit: Masked generative video transformer.
\newblock In \emph{Proceedings of the IEEE/CVF Conference on Computer Vision and Pattern Recognition}, pp.\  10459--10469, 2023.

\bibitem[Yu et~al.(2024{\natexlab{b}})Yu, Lezama, Gundavarapu, et~al.]{Yu2024LanguageModelBeats}
Luming Yu, Jose Lezama, Nikhil~B Gundavarapu, et~al.
\newblock Language model beats diffusion-tokenizer is key to visual generation.
\newblock In \emph{The Twelfth International Conference on Learning Representations}, 2024{\natexlab{b}}.

\bibitem[Yu et~al.(2022)Yu, Tack, Mo, Kim, Kim, Ha, and Shin]{yu2022generating}
Sihyun Yu, Jihoon Tack, Sangwoo Mo, Hyunsu Kim, Junho Kim, Jung-Woo Ha, and Jinwoo Shin.
\newblock Generating videos with dynamics-aware implicit generative adversarial networks.
\newblock In \emph{International Conference on Learning Representations}, 2022.

\bibitem[Zhang et~al.(2024)Zhang, Li, Le, Shou, Xiong, and Sahoo]{zhang2024moonshot}
David~Junhao Zhang, Dongxu Li, Hung Le, Mike~Zheng Shou, Caiming Xiong, and Doyen Sahoo.
\newblock Moonshot: Towards controllable video generation and editing with multimodal conditions.
\newblock \emph{arXiv preprint arXiv:2401.01827}, 2024.

\bibitem[Zhou et~al.(2024)Zhou, Huang, Xue, Dou, Cheng, Zhou, and Ni]{zhou2024heartbeat}
Xinrui Zhou, Yuhao Huang, Wufeng Xue, Haoran Dou, Jun Cheng, Han Zhou, and Dong Ni.
\newblock Heartbeat: Towards controllable echocardiography video synthesis with multimodal conditions-guided diffusion models.
\newblock \emph{arXiv preprint arXiv:2406.14098}, 2024.

\end{thebibliography}
\bibliographystyle{iclr2025_conference}

\newpage
\appendix
\section{Appendix}
\subsection{Reproducibility Statement}
To ensure the reproducibility of our results, we will make all relevant resources publicly available. This includes the complete codebase, training scripts, pre-trained model weights, and synthesized ECHO video data. By providing these materials, we encourage the research community to replicate and build upon our work. Detailed instructions on how to set up and run the experiments will also be included in our repository to facilitate easy reproduction of our results.
\subsection{MODEL SETUP AND HYPERPARAMETERS}
\subsubsection{Video Tokenizer}
\label{hyper}
The configuration of the video tokenizer model is as follows:

\begin{table}[htbp]
    \centering
    \begin{tabular}{|l|l|}
        \hline
        \textbf{Parameter} & \textbf{Value} \\ \hline
        Video input & 11 frames, frame stride 1, 128 $\times$ 128 resolution \\ \hline
        Batch size & 128 \\ \hline
        Embedding size & 512 \\ \hline
        Codebook size & 8192 \\ \hline
        Spatial patch size & 8 (non-overlapping) \\ \hline
        Temporal patch size & 2 (along the temporal axis) \\ \hline
        Spatial transformer depth & 4 layers \\ \hline
        Temporal transformer depth & 4 layers \\ \hline
        Transformer head dimension & 64 \\ \hline
        Number of heads & 8 \\ \hline
        Feed-forward multiplier & 4 (MLP size of 2048, i.e., $32 \times 64$) \\ \hline
    \end{tabular}
    \caption{Video tokenizer hyperparameters}
    \label{tab:model_hyperparameters}
\end{table}

\subsubsection{Video GENERATION TRANSFORMER}
The configuration of the transformer model in Figure \ref{pipeline}b is as follows:
\begin{table}[htbp]
    \centering
    \begin{tabular}{|l|l|}
        \hline
        \textbf{Parameter} & \textbf{Value} \\ \hline
        Number of tokens & 8192 (aligned with the codebook size) \\ \hline
        Batch size & 128 \\ \hline
        Embedding dimension & 512 \\ \hline
        ECG signal dimension & 768 \\ \hline
        Depth (number of layers) & 6 \\ \hline
        Learning rate & 1e-4 \\ \hline
        Training steps & 4,000,000 steps \\ \hline
        Exponential Moving Average (EMA) & Updated every 10 steps with a decay rate of 0.995 \\ \hline
        Optimizer & Adam optimizer with $\beta$ values set to (0.9, 0.99) \\ \hline
    \end{tabular}
    \caption{Video generation model hyperparameters}
    \label{tab:training_hyperparameters}
\end{table}

\subsection{Dataset}
\begin{figure}[ht]

\begin{center}
\includegraphics[width=\textwidth]{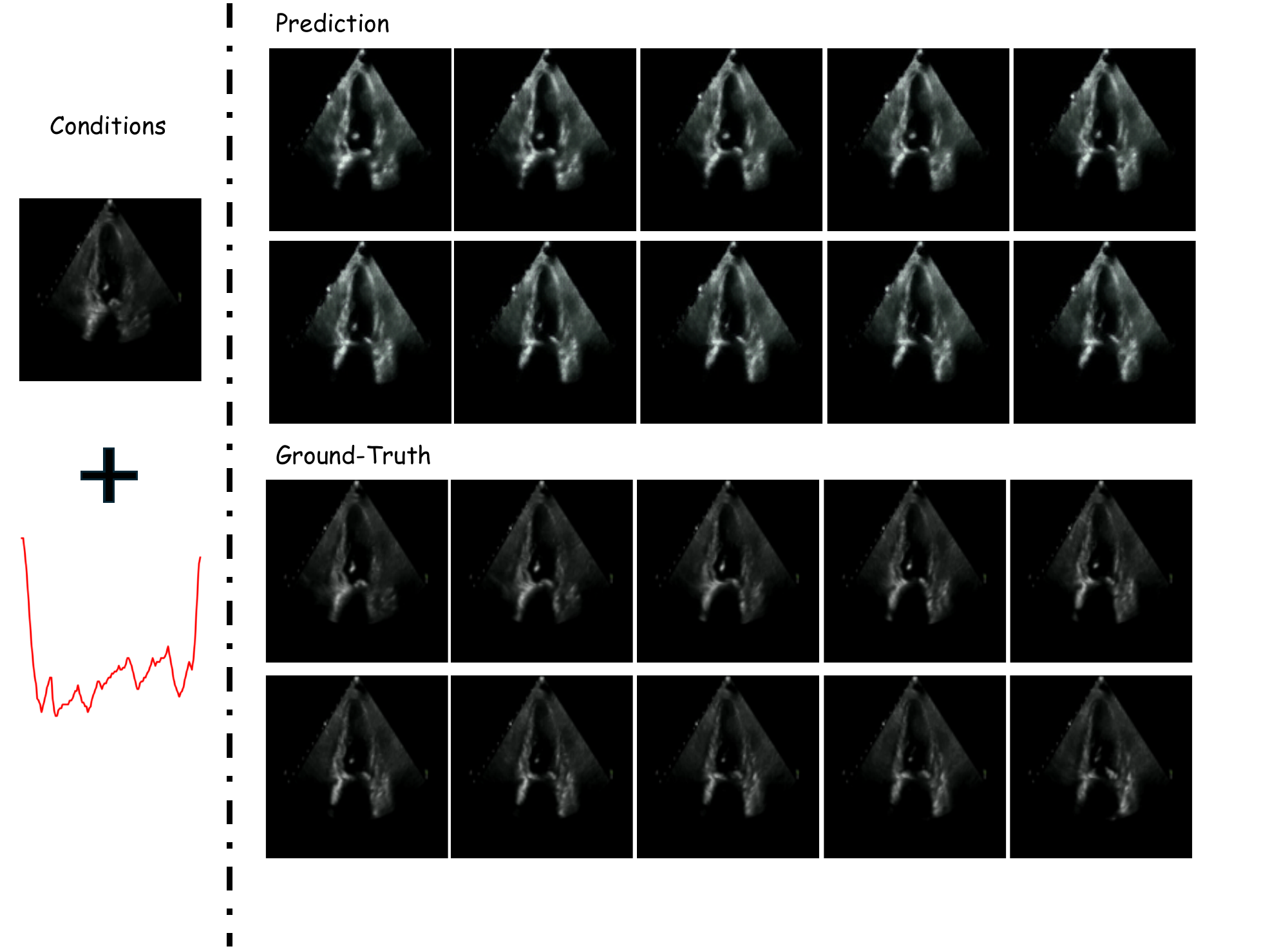}
\end{center}
\caption{A2C video generation example from \projectname{}.}
 \label{ap1}
\end{figure}

\begin{figure}[ht]
\begin{center}
\includegraphics[width=\textwidth]{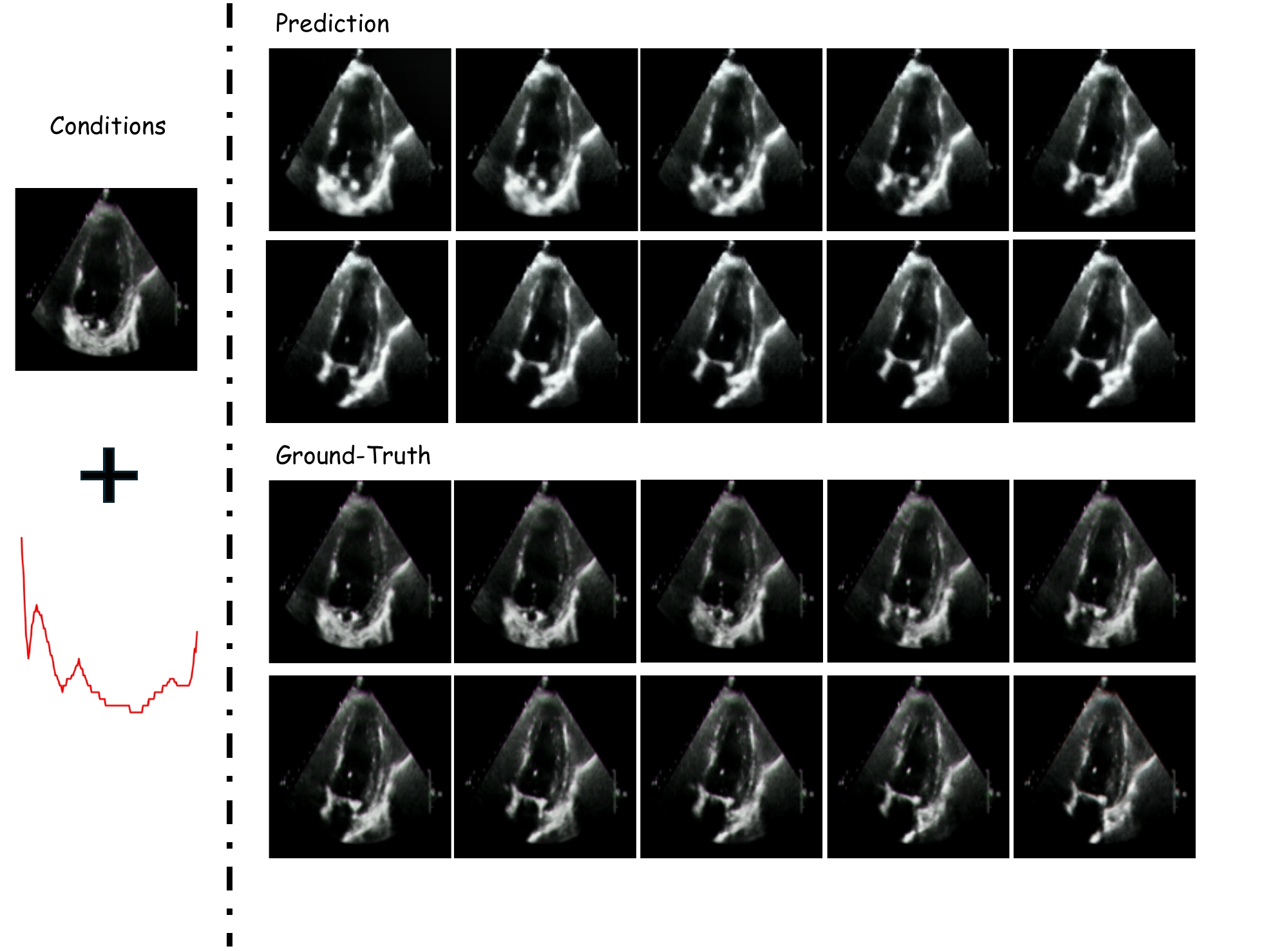}
\end{center}
\caption{A2C video generation example from \projectname{}.}
 \label{ap2}
\end{figure}

\begin{figure}[ht]
\begin{center}
\includegraphics[width=\textwidth]{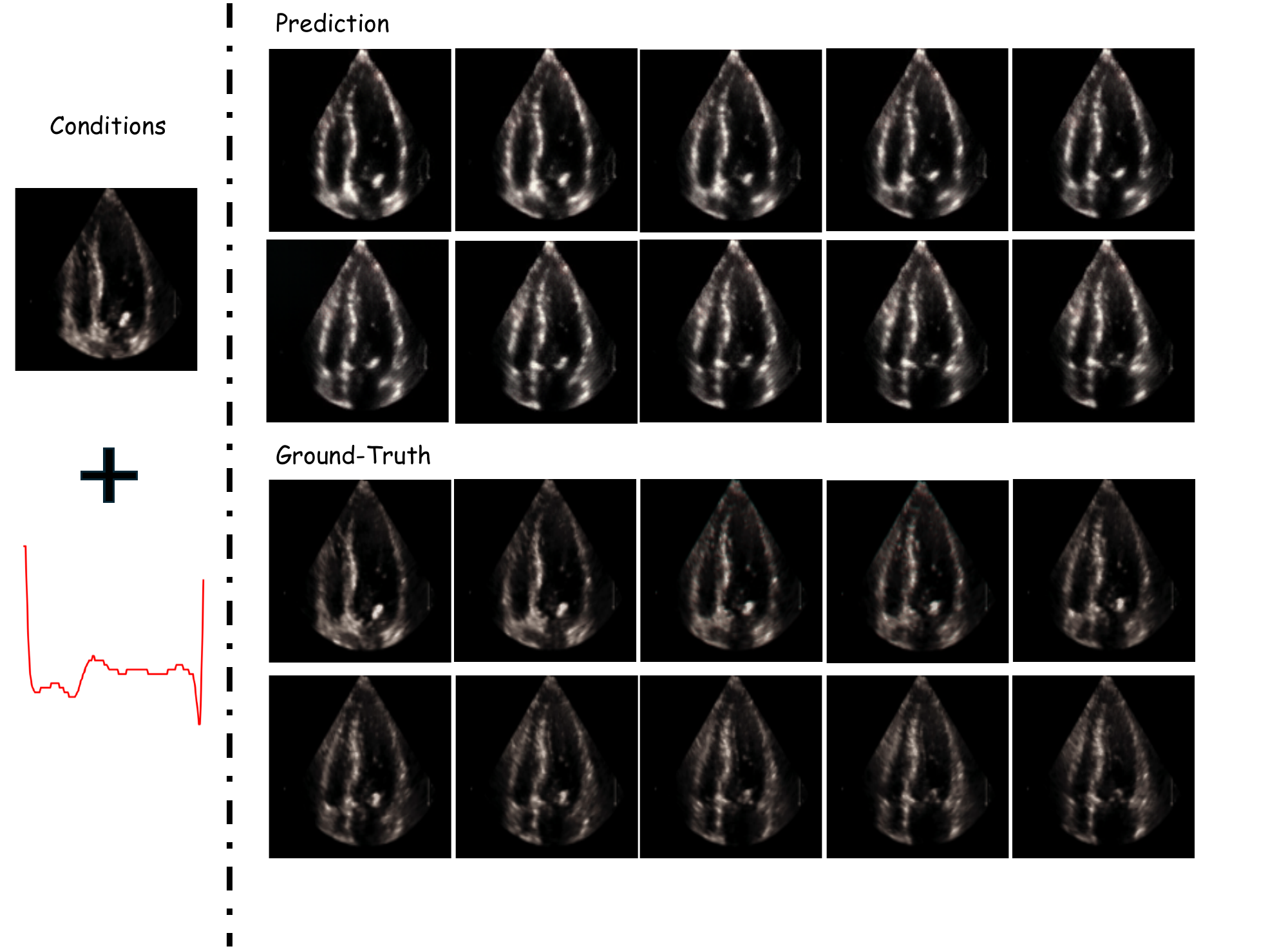}
\end{center}
\caption{A4C video generation example from \projectname{}.}
 \label{ap3}
\end{figure}

\begin{figure}[ht]
\begin{center}
\includegraphics[width=\textwidth]{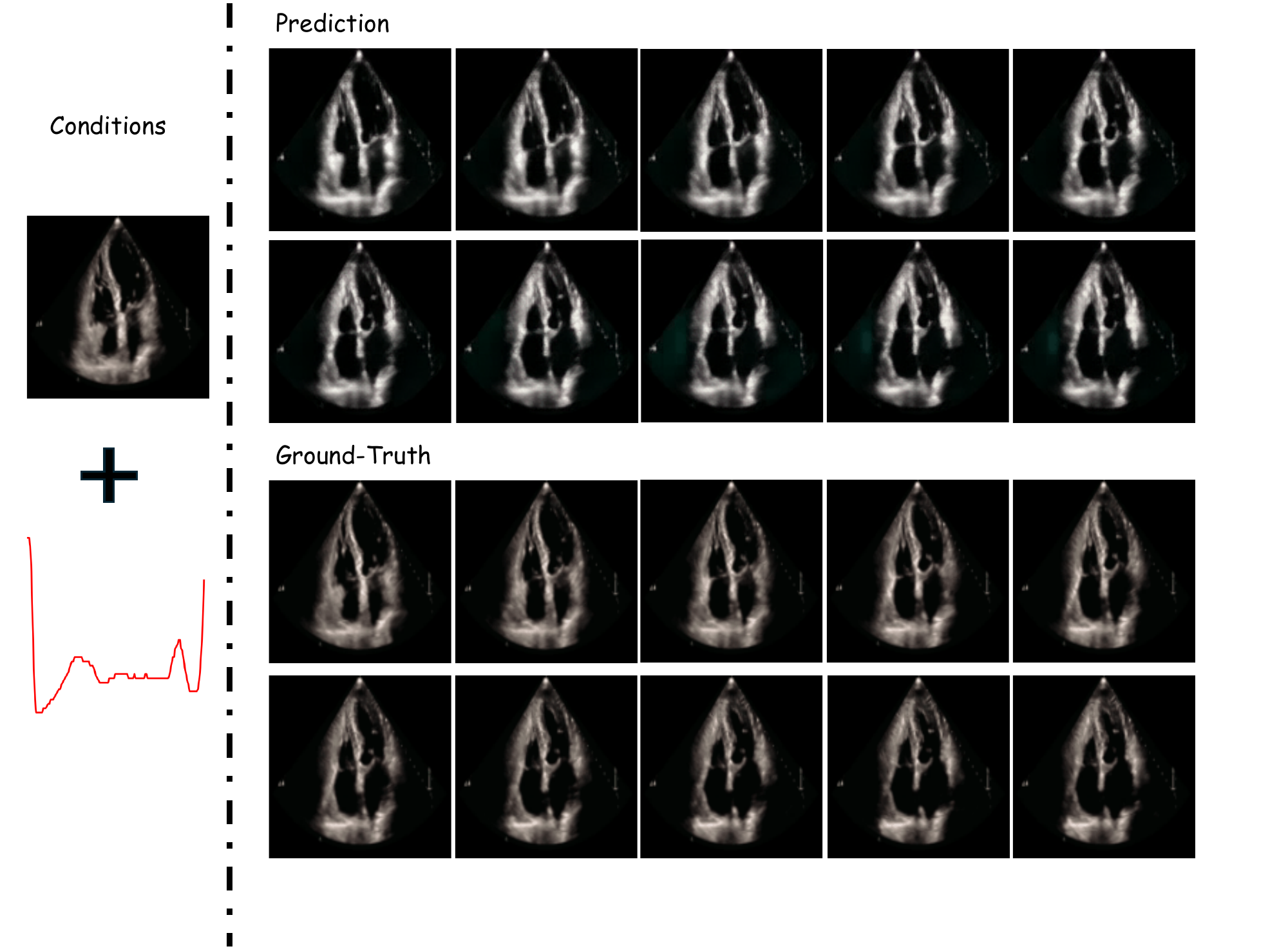}
\end{center}
\caption{A4C video generation example from \projectname{}.}
 \label{ap4}
\end{figure}

Our private dataset consists of 94,078 ECHO videos, including both apical 2 view and apical 4 view. It contains ECHO recordings from both healthy individuals and patients with various cardiovascular diseases, with real-time ECG signals recorded for each video. All videos were anonymized, underwent random contrast enhancement, and were resized to 128x128 resolution. Additional video generation examples are shown in the Figure \ref{ap1},\ref{ap2},\ref{ap3} and \ref{ap4}.

.

\end{document}